\documentstyle[12pt,epsf,epsfig]{article}     

\def\lsim{\lower0.6ex\vbox{\hbox{$ \buildrel{\textstyle 
<}\over{\sim}\ $}}}
\def\rsim{\lower0.6ex\vbox{\hbox{$ \buildrel{\textstyle 
>}\over{\sim}\ $}}}
\def\beq{\begin{equation}}
\def\eeq{\end{equation}}
\def\beginapjbib{\begingroup \section*{\large \bf References}
         \parskip=.5ex plus 1.0pt
         \def\bibitem{\par \noindent \hangindent\parindent
                \hangafter=1}}
\def\endapjbib{\par \endgroup}
\def\alwaysmath#1{{\ifmmode{#1}\else{$#1$}\fi}}
\def\he#1{\hbox{\alwaysmath{{}^{#1}}{\rm He}}}

\def\hii{H\thinspace{$\scriptstyle{\rm II}$}~}
\def\etal{{\it et al.}~}

\begin{document}
\begin{flushright}
OSU-TA-6/97  \\
April 1997\\
\end{flushright}
\vskip 0.7in
 
\begin{center} 

{\Large{\bf Temperature Fluctuations and Abundances in \hii Galaxies}}
 
\vskip 0.4in
{Gary Steigman$^1$, Sueli M. Viegas$^2$ and Ruth Gruenwald$^2$}
 
\vskip 0.2in
{\it $^1${Departments of Physics and Astronomy,
The Ohio State University, \\ 
Columbus, OH 43210, USA}\\

\vskip 0.1in 
$^2${Instituto Astron$\hat{o}$mico e Geof\' \i sico, 
Universidade de S$\tilde{a}$o Paulo, \\
S$\tilde{a}$o Paulo, S.P. 04301-904, BRASIL}\\}

\vskip .4in
{\bf Abstract}
\end{center}
\baselineskip=18pt 

 There is evidence for temperature fluctuations in Planetary 
Nebulae and in Galactic \hii regions.  If such fluctuations 
occur in the low-metallicity, extragalactic \hii regions used 
to probe the primordial helium abundance, the derived $^4$He 
mass fraction, Y$_{\rm P}$, could be systematically different 
from the true primordial value.  For cooler, mainly high-metallicity 
\hii regions the derived helium abundance may be nearly unchanged 
but the oxygen abundance could have been seriously underestimated.  
For hotter, mainly low-metallicity \hii regions the oxygen abundance 
is likely accurate but the helium abundance could be underestimated.  
The net effect is to tilt the Y vs. Z relation, making it flatter 
and resulting in a higher inferred Y$_{\rm P}$.  Although this effect 
could be large, there are no data which allow us to estimate the size 
of the temperature fluctuations for the extragalactic \hii regions.  
Therefore, we have explored this effect via Monte Carlos in which the
abundances derived from a fiducial data set are modified by $\Delta$T 
chosen from a distribution with 0 $\leq$ $\Delta$T $\leq$ $\Delta$T$_{max}$ 
where $\Delta$T$_{max}$ is varied from 500K to 4000K.  It is interesting 
that although this effect shifts the locations of the \hii regions in  Y vs. 
O/H plane, it does not introduce any significant additional dispersion.

\newpage
\baselineskip=18pt 
\noindent
\section{Introduction}

The primordial abundance of \he4 is key to testing the consistency of 
the standard hot big bang model of cosmology and to using primordial
nucleosynthesis as a probe of particle physics (Steigman, Schramm \& 
Gunn 1977).  The availability of large numbers of carefully observed, 
low-metallicity \hii regions has permitted estimates of the primordial
helium mass fraction, $Y_{\rm P}$, whose statistical uncertainties are 
very small, $\approx$ 1\% (see, Olive \& Steigman 1995 (OS), Olive, 
Skillman \& Steigman 1997 (OSS) and references therein).  However, there 
remains the possibility that in the process of using the observational 
data to derive the abundances, contamination by unacknowledged systematic 
uncertainties has biased the inferred value of $Y_{\rm P}$.  Observers 
have identified many potential sources of systematic uncertainties 
(Davidson \& Kinman 1985; Pagel \etal 1992 (PSTE); Skillman \etal 1994; 
Izotov, Thuan \& Lipovetsky 1994, 1997 (ITL); Peimbert 1996) and, where 
possible, have designed their observing programs to minimize such 
uncertainties and/or to account for them.  It is expected that the 
contributions from many of the potential sources of systematic uncertainty 
(if present at all) would vary from \hii region to \hii region and from 
observer (telescope/detector combination) to observer, introducing along 
with a systematic offset in the derived value of $Y_{\rm P}$, an 
accompanying dispersion in the helium data.  More insidious would 
be systematic errors in, for example, the atomic physics used to 
convert the observed equivalent widths to abundances, since an offset
from the ``true" value of $Y_{\rm P}$ would not be accompanied by an
enhanced dispersion in the fits to the data (e.g., Y vs. O/H or the
weighted mean of Y for the lowest metallicity \hii regions; see OSS).
In this paper we explore one potential source of systematic uncertainty
in deriving abundances from emission-line data for extragalactic, \hii
regions: temperature fluctuations.  

The empirical method to derive chemical abundances in an
emitting gas has been used widely since it was first proposed 
by Peimbert and Costero (1969). The electron density n$_e$ 
and the gas temperature T are obtained from emission-line 
intensity ratios and are used to calculate the line-emissivity
which, along with the observed line-intensities, provide the 
fractional abundance of the ion and, subsequently, the empirical 
abundance of a given element  (see, for example, Osterbrock 1989). 
In  low density regions, such as those in \hii galaxies, the
permitted-line emissivities (like those of H and He) decrease 
(slowly) with T and are usually independent of n$_e$, unless 
collisional effects are important.  On the other hand, the 
forbidden-line emissivities (like those of O, N, S) are strongly 
dependent on T, and the dependence on the density may also be 
important.  In \hii galaxies, the [S II] line ratio indicates 
low electron densities (n$_e$ $<$ 500 cm$^{-3}$), and except for 
those with high temperature, the collisional effects are negligible. 
Thus, the derived heavy element abundances depend mainly on a good 
determination of the temperature.  For these regions, the temperature 
used for the high-ionization lines is that obtained from the [O III] 
line ratio, T$_{OIII}$, while for the low-ionization lines the 
temperature is derived from T$_{OIII}$ using results from photoionization 
models (see for example, PSTE). 

A similar method is usually applied to planetary nebulae.
In contrast to \hii regions, for several PNe the Balmer 
temperature T$_B$ (obtained from the observed Balmer 
discontinuity) is also determined.  In many cases T$_B$ 
is found to be lower than T$_{OIII}$ (Peimbert 1971; Liu 
\& Danziger 1993).  As pointed out by Peimbert (1971), 
this discrepancy could be due to temperature fluctuations, 
which, however, are not reproduced by the standard photoionization 
models for these nebulae (Liu \& Danziger 1993).  If, indeed, 
the gas temperature is overestimated by T$_{OIII}$, with the 
true temperature given by T$_B$, the heavy element abundances 
derived from forbidden lines using T$_{OIII}$ are underestimated 
(Viegas \& Clegg 1994).  Indeed, the abnormal chemical abundances 
inferred for some \hii galaxies with WR features may result from 
just such an overestimate of the gas temperature (Esteban \& Peimbert 
1995).  A well-observed giant \hii region, NGC 2363, shows just such 
a discrepancy between the Paschen temperature and T$_{OIII}$ in two 
knots (Gonz\'alez-Delgado et al. 1994).  For the \hii galaxies used 
to derive the primordial helium abundance, such an underestimate of 
the true oxygen abundance may introduce a systematic bias into the 
inferred value of Y$_{\rm P}$.  Furthermore, temperature fluctuations 
also will have a direct effect on the helium abundance determined 
from the recombination lines (Esteban \& Peimbert 1995).
Although the ratio of He and H emissivities is very nearly independent
of temperature, the helium-line intensities are modified from the pure
recombination case by collisional excitation (Cox \& Daltabuit 1971;
Ferland 1986) which is temperature-sensitive.  If the ``true" \hii 
region temperature has been overestimated, the collisional correction 
has been overestimated and the ``true" helium abundance underestimated.  
This effect will be largest for the hottest, metal-poor \hii regions.
Thus the combined effects, which tend to increase the derived oxygen
abundance in the metal-rich \hii regions and the helium abundance
in the metal-poor regions (see Figure 1), will ``tilt" the inferred Y 
versus O/H relation, flattening the slope and increasing the intercept, 
Y$_{\rm P}$.  In this paper we analyse these effects in an attempt to 
quantify the corresponding systematic uncertainty in Y$_{\rm P}$.

\section{The Method}

If the Paschen or Balmer temperatures were available, the corrections
described above would be straightforward to implement.  Unfortunately, 
for the low-metallicity, extragalactic \hii regions used to derive the 
primordial abundance of helium, there appears to be no data on the Balmer 
or Paschen temperatures.  To estimate the magnitude of the possible
corrections we have, therefore, adopted a Monte Carlo approach to 
quantifying the potential systematic uncertainty introduced into the 
determination of Y$_{\rm P}$ by uncertainty in the temperature of 
the \hii region gas.

We begin by adopting a ``fiducial" data set using 43 \hii galaxies 
from PSTE, ITL, Skillman \& Kennicutt (1993) and Skillman \etal (1994).  
The role of this fiducial set is merely to provide a comparison, in 
order to quantify the {\it changes} induced by temperature differences.  
For this reason we have not employed the unpublished data of Skillman 
et al. (in preparation, see OS and OSS) nor those \hii regions listed 
in PSTE which they, themselves, did not observe.  We follow ITL and 
eliminate those regions for which their data is noisy or otherwise 
uncertain.  For each \hii region in our fiducial data set we adopt the 
same algorithms used by these authors to derive the temperature and the 
fractional abundances in the low-ionization zones.  The electron density 
is obtained from the [S II] line ratio and we use the Brocklehurst 
(1972) recombination coefficients for the He and H lines along with the 
collisional correction for the He lines from Clegg (1987).  The fractional 
abundance of He$^+$ is the (unweighted) average of the abundances obtained 
from HeI $\lambda$4471, HeI $\lambda$5876 and HeI $\lambda$6678.  
Ignoring any statistical uncertainties, we use these fiducial 
abundances to find a ``standard" linear fit to Y versus O/H 
(unweighted) in the absence of temperature fluctuations.  We emphasize 
that we are not so much interested in the ``best fit" Y versus O/H relation 
for our fiducial data set but, rather, the {\it differences} in the Y versus 
O/H relations between $\Delta$T zero and nonzero ($\Delta$T $\equiv$ 
T$_{OIII} - $T).  In Figure 1 each point in our data set is shown in the 
Y versus O/H plane in the absence of temperature fluctuations ($\Delta$T 
= 0) and for $\Delta$T = 2000K, joined by a solid line.  As anticipated, 
the higher metallicity regions tend to be cooler (see Figure 2).  This 
has the effect that the derived oxygen abundance in high-metallicity 
regions is very sensitive to temperature fluctuations as is seen clearly 
in Figure 1.

For planetary nebulae the observed temperature difference $\Delta$T =
T$_{OIII} - $T$_B$ is typically 2000K (Liu \& Danziger 1993) with
one PN having $\Delta$T = 6000K.  For the giant \hii region NGC 2363, 
$\Delta$T $\geq$ 3000K (Gonz\'alez-Delgado et al. 1994).  Thus, we 
have recalculated the helium and oxygen abundances for each \hii region 
with $\Delta$T chosen from a distribution which ranges from zero to 
$\Delta$T$_{max}$, according to a probability P($\Delta$T) to be described
below; $\Delta$T$_{max}$ is varied in the range 500 $\leq \Delta$T$_{max}$ 
$\leq$ 4000K.  We repeat this procedure 10,000 times for each choice of 
$\Delta$T$_{max}$.  For most \hii regions Y does not decrease with decreasing 
temperature as might have been expected from the effective recombination 
coefficients alone (Esteban \& Peimbert 1995) provided we account for the 
collisional effect on the He lines; for \hii galaxies with high gas 
temperature, the collisional effect dominates, leading to an increase of Y.  
As seen in Figure 1, the decrease in T leads to an increase in the oxygen 
abundance.  For each of the 10,000 realizations (for each choice of
$\Delta$T$_{max}$) of our \hii region data set we fit a linear Y versus 
O/H relation to derive Y$_{\rm P}$ and the slope $\Delta$Y/$\Delta$Z, 
where the heavy element mass fraction Z $\approx$ 20(O/H) (see PSTE).

\section{Results: Y Versus Z}

\subsection{Full Data Set}

The results for our full data set of 43 \hii regions, selected to have
O/H $\leq$ 1.5$\times$ 10$^{-4}$, are shown in Figures 3 -- 5.  As 
anticipated, the intercept, Y$_{\rm P}$, is systematically increased by 
an amount which depends on $\Delta$T$_{max}$ (see Figures 3a \& 5).  This
increase in the intercept of the Y versus Z relation is accompanied by a
flattening of the slope (see Figures 3b \& 5).  To probe the sensitivity
of our results to the adopted probability distribution of $\Delta$T values, 
P($\Delta$T), we have considered three, simple distributions: flat, linearly 
increasing, linearly decreasing.  The results for these three, different 
distributions are shown for $\Delta$T$_{max}$ = 4000K in Figure 4.  Clearly, 
the effects are closely similar for all choices; in our subsequent discussion 
(and, in Figures 3 \& 5) we present results for the flat distribution 
(P($\Delta$T) independent of $\Delta$T).

In Figure 5 is shown how the magnitude of the systematic offsets in slope
and intercept scale with $\Delta$T$_{max}$.  For the not unreasonable 
choice of $\Delta$T$_{max}$ = 2000K, the increase in the inferred value 
of Y$_{\rm P}$ is significant, comparable to some estimates of the upper 
bound to the systematic uncertainty in Y (OS; OSS).  For larger values of 
$\Delta$T$_{max}$, the systematic offset will be even larger.  It might 
have been anticipated (see OSS) that such large systematic offsets would 
be accompanied by an increase in the dispersion of the abundances around 
the best fit Y versus Z relation.  To test for this, for each realization 
we have computed $\sigma$, the variance of the residuals between the data 
(Y) and those values predicted by the corresponding fit for that realization 
(Y$_{fit}$).  However, as seen in Figure 5, for temperature fluctuations 
this effect of added dispersion, if present at all, is very small compared, 
for example, to the typical uncertainty in the individual \hii region Y 
determinations which are of order 0.010 (OSS).  Thus, until there are 
observations of \hii region temperatures determined from hydrogen lines, 
it is difficult to set a firm upper bound to the magnitude of the 
systematic offset in Y$_{\rm P}$ due to the possible presence of 
temperature fluctuations.  It is, therefore, important to consider 
how best to analyze current data so as to minimize the potential 
importance of such temperature fluctuations.

\subsection{Truncated Data Set}

A significant contribution to the systematic offset in Y$_{\rm P}$ in 
the presence of temperature fluctuations comes from the flattening of 
the slope of the Y versus Z relation due to the large increase in oxygen 
abundance for the cooler, higher-metallicity \hii regions (see Figures 1 
\& 2).  Therefore, the uncertainty in Y$_{\rm P}$ derived from a fit to 
the trend of Y with O/H might be minimized if the data set is restricted 
to the very lowest metal abundance \hii regions which are hotter.  To 
explore this, for each realization of our Monte Carlos we identify the 
subset of \hii regions with low-metallicity: O/H $\leq$ 0.9$\times$ 10$^{-4}$.  
For these subsets we fit linear Y versus Z relations and compare the slopes 
and intercepts (as well as $\sigma$) to those found for the corresponding
low-metallicity subset in the absence of temperature fluctuations.  The 
results are shown in Figures 6 \& 7.  As expected, the changes in slope 
and intercept for this low-metallicity set are much smaller.  Indeed, the 
trend is even opposite that for our full data set: {\it lower} intercept, 
{\it higher} slope (although the effect is so small as to be only marginally 
significant).  The lesson is clear.  If we wish to minimize the uncertainty 
due to temperature fluctuations in Y$_{\rm P}$ derived from a fit of Y 
versus Z, we should restrict our attention to the most metal-poor \hii regions.

\subsection{Lowest-Y \hii Regions}

Since it is generally accepted that the helium abundance has only increased
since the big bang, an alternate approach to using \hii regions observations
to infer the primordial abundance is to compute the mean of Y for those
regions with the lowest helium abundances; then, Y$_{\rm P}$ $\leq$ $<$Y$>$
(OS; OSS).  This estimator is likely to be robust against the systematic
shifts due to temperature fluctuations.  To test this we have first chosen
the ten lowest-Y \hii regions in our fiducial data set ($\Delta$T = 0) and 
found the unweighted average of Y, $<$Y$>_{10}$(0).  Then, from our Monte 
Carlos, for various choices of $\Delta$T$_{max}$ we determine (for each 
realization) the ten lowest-Y \hii regions (often, but not always, the same
as in the fiducial set) and compute $\Delta$$<$Y$>_{10}$ $\equiv$  
$<$Y$>_{10}$($\Delta$T) $- <$Y$>_{10}$(0).  The distributions of 
$\Delta$$<$Y$>_{10}$ for $\Delta$T$_{max}$ = 2000K and 4000K are shown in
Figure 8 and the trend of $\Delta$$<$Y$>_{10}$ with $\Delta$T$_{max}$
is shown in Figure 9 where the corresponding change in the dispersion 
around the mean ($\sigma$) is also shown.  As expected, although temperature 
fluctuations tend to increase $<$Y$>_{10}$ systematically (mainly due to 
the reduced correction for collisional excitation), the effect is quite small.

\section{Discussion}

Temperature fluctuations in the low-metallicity, extragalactic \hii regions
used to infer the primordial helium abundance will lead to systematic offsets
in the helium and oxygen abundances derived for those regions.  In the absence
of direct observations of the hydrogen temperatures for these regions we have
attempted to quantify the effect of such fluctuations on the derived value of
Y$_{\rm P}$ by performing a series of Monte Carlo calculations where the
temperature fluctuation for each \hii region is chosen from a distribution
whose maximum value is varied.  Y$_{\rm P}$ determined from a linear fit to
Y versus O/H is especially susceptible to this source of error which flattens
the Y versus Z relation resulting in a higher intercept.  This potential
systematic error is insidious in the sense that the offset in Y$_{\rm P}$ 
may be significant without a noticeable increase in the dispersion of the
individual \hii region Y-values around the best fit Y versus Z relation.
To illustrate the possible offsets which may be consistent with extant 
data, consider the results of OSS for their full data set of 62 \hii 
regions with O/H $\leq$ 1.5$\times10^{-4}$: Y$_{\rm P}$ = 0.234$\pm$0.005 
(0.005 is a 2-$\sigma$ estimate of the statistical uncertainty).  For 
$\Delta$T$_{max}$ = 4000K we should increase this value by 0.008$\pm$0.004 
(see Figure 5).  Adding the (uncorrelated) systematic and statistical 
uncertainties in quadrature will lead us to Y$_{\rm P}$ = 0.242$\pm$0.006 
(95\%CL).  Such a large correction to Y$_{\rm P}$ would ameliorate the 
``tension" in BBN (Hata et al. 1995) between primordial helium and the 
low abundance of deuterium suggested by Galactic observations (see, e.g., 
Dearborn, Steigman \& Tosi 1995 and references therein) and by some 
direct detections of deuterium in low-metallicity, high-redshift QSO 
absorbers (Tytler, Fan \& Burles 1996; Burles \& Tytler 1996).

Until data on temperature fluctuations in these key extragalactic
\hii regions become available, the best strategy is to analyze the 
current data in a manner which minimizes this potential systematic error.
One possibility is to restrict attention to the lowest-metallicity regions
available.  OS and OSS have done this for the subset with O/H $\leq$ 0.9
$\times10^{-4}$ for which OSS derive from a linear fit to Y versus O/H:
Y$_{\rm P}$ = 0.230$\pm$0.007 (95\%CL).  From our Monte Carlos we find for 
this subset that the systematic correction is likely very small (indeed, it 
may even be negative!); for $\Delta$T$_{max}$ = 4000K, $\Delta$Y$_{\rm P}$ 
= $-0.004\pm$0.005 (see Figure 7).  If, perhaps naively, we apply this 
correction to the OSS result, we infer: Y$_{\rm P}$ = 0.226$\pm$0.009 
(95\%CL).  In this case the ``tension" between primordial helium and 
low primordial deuterium is exacerbated.

Another ``safe" approach to using the existing data to derive an estimate 
of Y$_{\rm P}$ is to ignore the metallicity information and simply take a
mean ($<$Y$>$) of the helium abundance for the lowest abundance \hii regions.  
Since helium is only expected to increase after BBN, this provides an upper 
bound to Y$_{\rm P}$.  In general, the \hii regions with the lowest values 
of Y tend to be the lowest-metallicity regions which are also the hottest 
(see Figure 2).  For such regions there has often been a non-negligible 
correction for collisions in deriving Y from the helium line intensities.  
If the gas is, in fact, cooler, this correction has been overestimated and 
the ``true" value of Y should be larger.  Thus, temperature fluctuations 
will increase $<$Y$>$.  From our Monte Carlos we have selected, for each
realization, the ten \hii regions with the lowest Y-values and we have
found, for $\Delta$T$_{max}$ = 4000K, a small systematic increase: 
$\Delta$$<$Y$>_{10}$ = 0.002$\pm$0.001 (see Figures 8 \& 9).  For their 
ten lowest-Y regions, OSS find $<$Y$>_{10}$ = 0.230$\pm$0.006 (95\%CL), 
so that even with our largest correction we infer a revised mean 
of 0.232$\pm$0.006 suggesting that Y$_{\rm P} \leq$ 0.238 (95\%CL).  
Here, too, the ``tension" between D and $^4$He fails to be relieved.

\section{Summary}

Temperature fluctuations in low-metallicity, extragalactic \hii regions 
may have a significant effect on the determination of the primordial
abundance of helium.  If present, they may increase the metallicity of
the higher-metallicity, relatively cooler regions and increase the
helium abundance of the more metal-poor, hotter regions, tilting the
inferred Y versus Z relation and leading to a higher, zero-metallicity
intercept (Y$_{\rm P}$).  Such a systematic offset is not necessarily
accompanied by a significant increase in the dispersion of the data
around the best fit linear Y versus Z relation and, therefore, may remain
invisible in the absence of direct data on temperature differences in
these regions.  It is clear that such data is crucial to constraining
the uncertainty in Y$_{\rm P}$.  In the absence of such data, we have
noted that restricting attention to the lowest-metallicity regions (Z
$\leq$ Z$_{\odot}$/10) will tend to minimize this systematic offset.
Alternatively, the mean of the helium abundances of the lowest-Y \hii
regions is also robust against the effect of temperature fluctuations.

\vskip 0.5truecm

\noindent {\bf Acknowledgments}

\vskip 0.5truecm

S.M.V. and R.G. owe a debt of gratitude to P. Benevides-Soares for
valuable discussions.  The work of S.M.V. and R.G. is partially
supported by FAPESP and CNPq in Brasil; that of G.S. is supported
by the DOE.  This work was initiated when S.M.V. and G.S. were
visiting the INT at the University of Washington and they are grateful
for the hospitality and stimulating environment provided there.


\vskip 0.5truecm

\beginapjbib

\bibitem Brocklehurst, M. 1972, MNRAS, 157, 211

\bibitem Burles, S. \& Tytler, D. 1996, ApJ, 460, 584

\bibitem Clegg, R.E.S. 1987, MNRAS, 229, 31P

\bibitem Cox, D.P. \& Daltabuit, E. 1971, ApJ, 167, 257

\bibitem Davidson, K. \& Kinman, T.D. 1985, ApJS, 58, 321

\bibitem Esteban, C. \& Peimbert, M. 1995, A\&A, 300, 78

\bibitem Ferland, G.J. 1986, ApJ, 310, L67

\bibitem Gonz\'alez-Delgado, R.M., P\'erez, E., Tenorio-Tagle, G., 
Vilchez, J. M., Terlevich, E., Terlevich, R., Telles, E., Rodriguez-Espinosa,
J. M., Mas-Hesse, M., Garc\' \i a-Vargas, M. L., D\' \i az, A. I.,
Cepa, J. \& Castan\~eda , J. 1994, ApJ 437, 239

\bibitem Hata, N., Scherrer, R.J., Steigman, G., Thomas, D., Walker, T.P., 
Bludman, S. \& Langacker, P. 1995,  Phys. Rev. Lett., 75, 3977

\bibitem Izotov, Y.I., Thuan, T.X., \& Lipovetsky, V.A. 1994
ApJ 435, 647. (ITL)
 
\bibitem Izotov, Y.I., Thuan, T.X., \& Lipovetsky, V.A. 1997, 
ApJS, 108, 1  (ITL)

\bibitem Liu, X. \& Danziger, J. 1993, MNRAS, 263, 256

\bibitem Olive, K.A., Skillman, E. \& Steigman, G. 1997, ApJ, In Press (OSS)

\bibitem Olive, K.A., \& Steigman, G. 1995, ApJS, 97, 49 (OS)

\bibitem Osterbrock, D.E. 1989, Astrophysics of Gaseous Nebulae and 
   Active Galactic Nuclei, University Science Books, Mill Valley, California

\bibitem Pagel, B E.J., Simonson, E.A., Terlevich, R.J.
\& Edmunds, M. 1992, MNRAS, 255, 325 (PSTE)

\bibitem Peimbert, M. 1971, Bol. Obs. Tonantzintla y Tacubaya, 6, 29

\bibitem Peimbert, M. 1996, Rev. Mex. Astr. Astrofis., Serie de Conferencias, 
4, 55 

\bibitem Peimbert, M. \& Costero, R. 1969, Bol. Obs. Tonantzintla y Tacubaya, 
5, 3

\bibitem Skillman, E., \& Kennicutt, R.C. 1993, ApJ, 411, 655 

\bibitem Skillman, E., Terlevich, R.J., Kennicutt, R.C., Garnett, D.R., 
\& Terlevich, E. 1994, ApJ, 431, 172

\bibitem Steigman, G., Schramm, D.N. \& Gunn, J. 1977, Phys. Lett, B66, 202

\bibitem Tytler, D., Fan, X.-M., and Burles, S. 1996, Nature, 381, 207

\bibitem Viegas, S.M. \& Clegg, R. 1994, MNRAS, 271, 993

\endapjbib

\newpage
\noindent{\bf{Figure Captions}}

\vskip.3truein

\begin{itemize}

\item[]
\begin{enumerate}
\item[]
\begin{enumerate}

\item[{\bf  Figure 1:}] The effect of temperature fluctuations on the 
helium and oxygen abundances for our fiducial data set.  The left tick 
mark is for no temperature fluctuations ($\Delta$T = 0) and the right 
tick mark is for $\Delta$T = 2000K.

\item[{\bf Figure 2:}] The temperatures of the \hii regions in our fiducial
data set as a function of the oxygen abundance (evaluated for $\Delta$T = 0).

\item[{\bf Figure 3a:}] The distribution of offsets in the primordial
mass fraction of $^4$He (inferred from a linear fit to Y versus O/H) due
to temperature fluctuations for $\Delta$T$_{max}$ = 2000K (shaded histogram) 
and $\Delta$T$_{max}$ = 4000K.  The results here are for the full data set 
(O/H $\leq$ 1.5$\times 10^{-4}$).

\item[{\bf Figure 3b:}] The corresponding distribution of the ratios of 
the slopes of the Y versus Z (Z $\approx$ 20(O/H)) relations for 
$\Delta$T$_{max}$ = 2000K (shaded histogram) and $\Delta$T$_{max}$ = 
4000K compared to the fiducial slope (for $\Delta$T = 0).

\item[{\bf Figure 4:}] The distribution of offsets in the primordial
mass fraction of $^4$He evaluated for the full data set with 
$\Delta$T$_{max}$ = 4000K for three different temperature probability 
distributions (from top to bottom: flat, linearly increasing, linearly 
decreasing).  

\item[{\bf Figure 5:}] In the three panels the solid curves show the
variation of the mean values of the offsets in Y$_{\rm P}$ (top panel), 
the ratio of slopes (middle panel) and the dispersion around the best
fit linear Y vs. Z relation (bottom panel) as a function of $\Delta$T$_{max}$.
The dotted curves show the 95\%CL ranges.

\item[{\bf Figure 6a:}] Similar to Figure 3a, but for the low-metallicity
subset of \hii regions (O/H $\leq$ 0.9$\times 10^{-4}$).

\item[{\bf Figure 6b:}] Similar to Figure 3b, but for the low-metallicity
subset of \hii regions.

\item[{\bf Figure 7:}] Similar to Figure 5, but for the low-metallicity
subset of \hii regions (O/H $\leq$ 0.9$\times 10^{-4}$).

\item[{\bf Figure 8:}] The distribution of offsets in the means of the
ten lowest-Y \hii region helium abundances for $\Delta$T$_{max}$ = 2000K 
(shaded histogram) and  $\Delta$T$_{max}$ = 4000K.

\item[{\bf Figure 9:}] Similar to Figures 5 \& 7 for the means of the
ten lowest-Y \hii region helium abundances as a function of $\Delta$T$_{max}$.

\end{enumerate}
\end{enumerate}
\end{itemize}

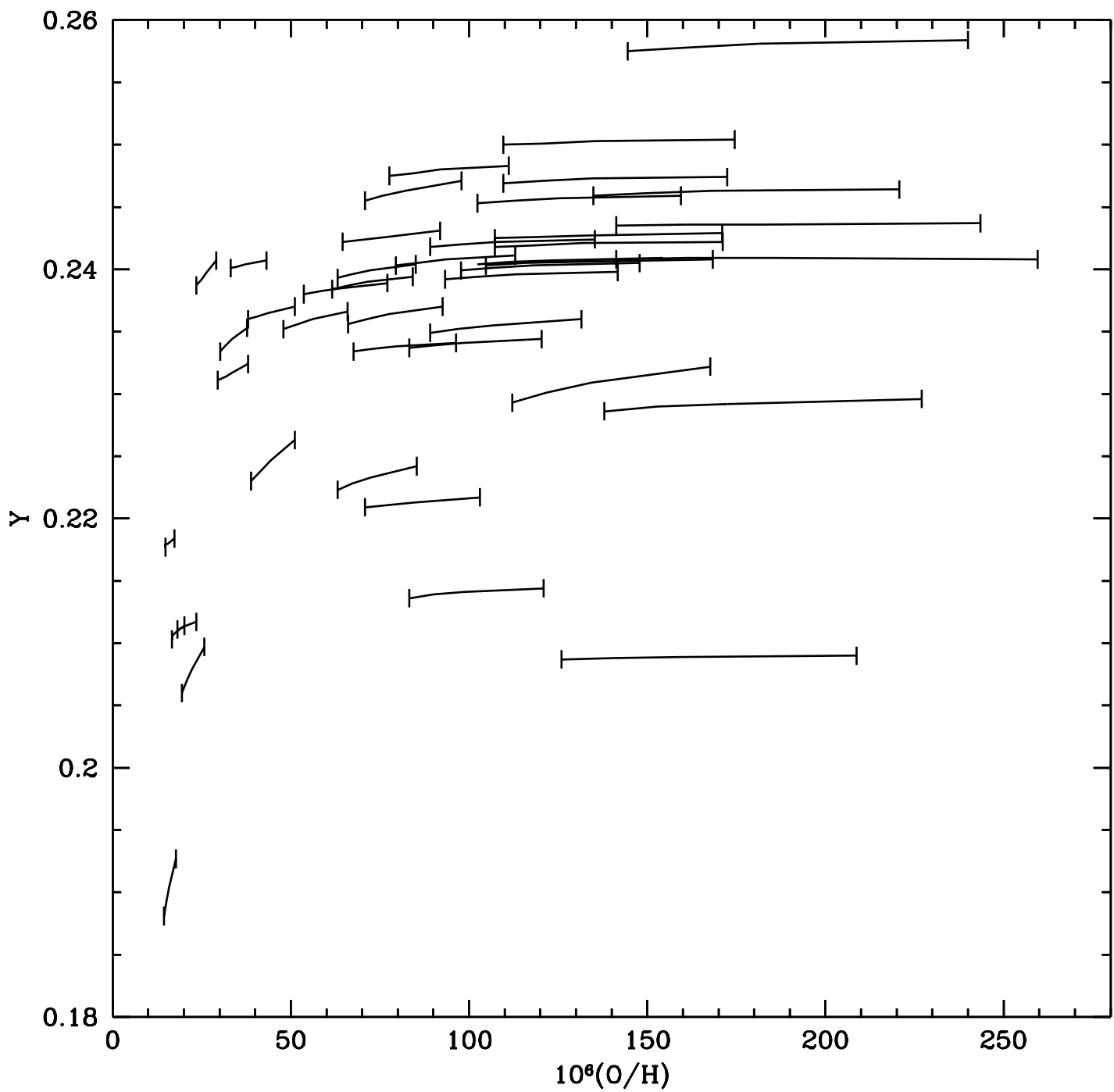
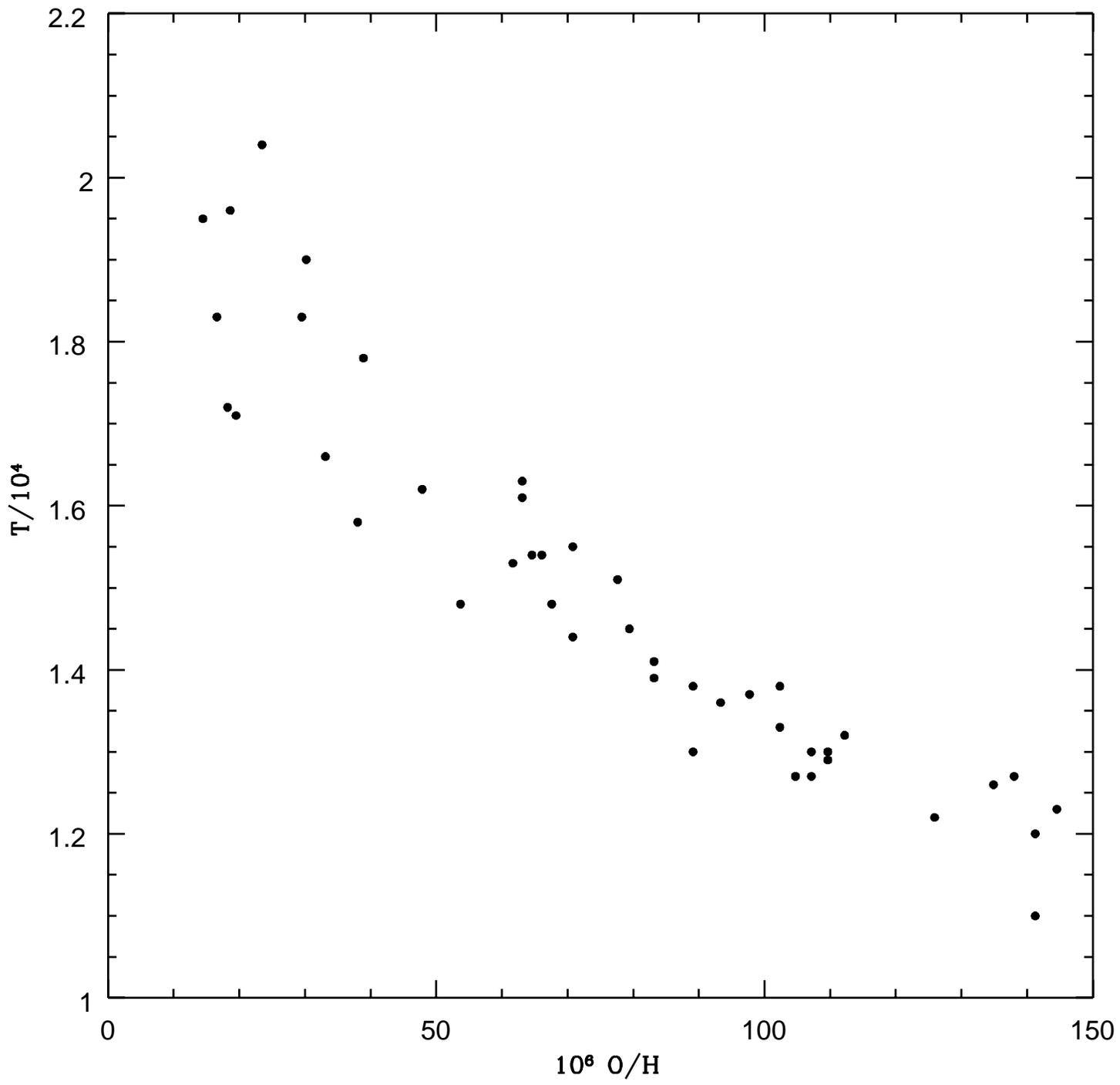
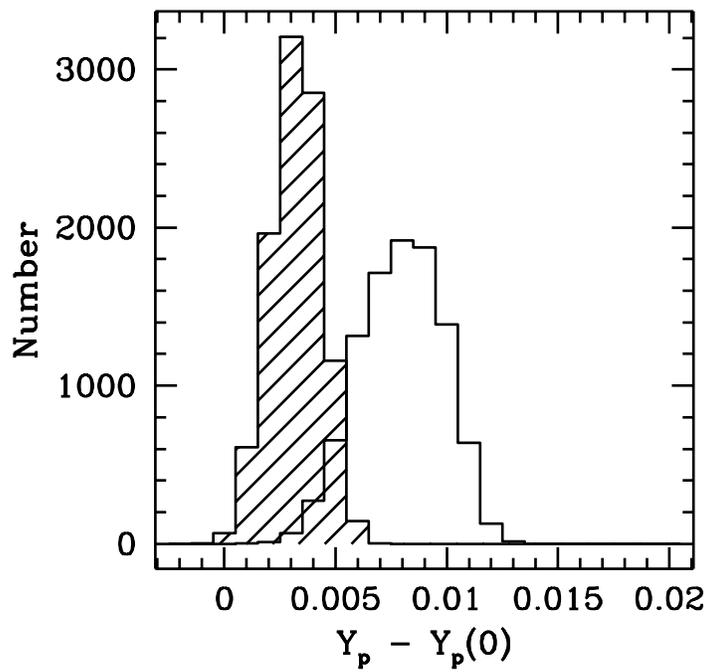
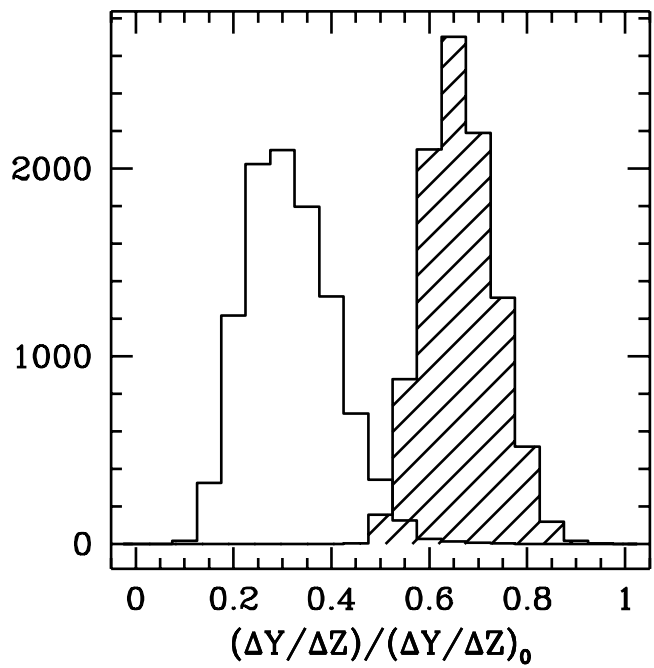
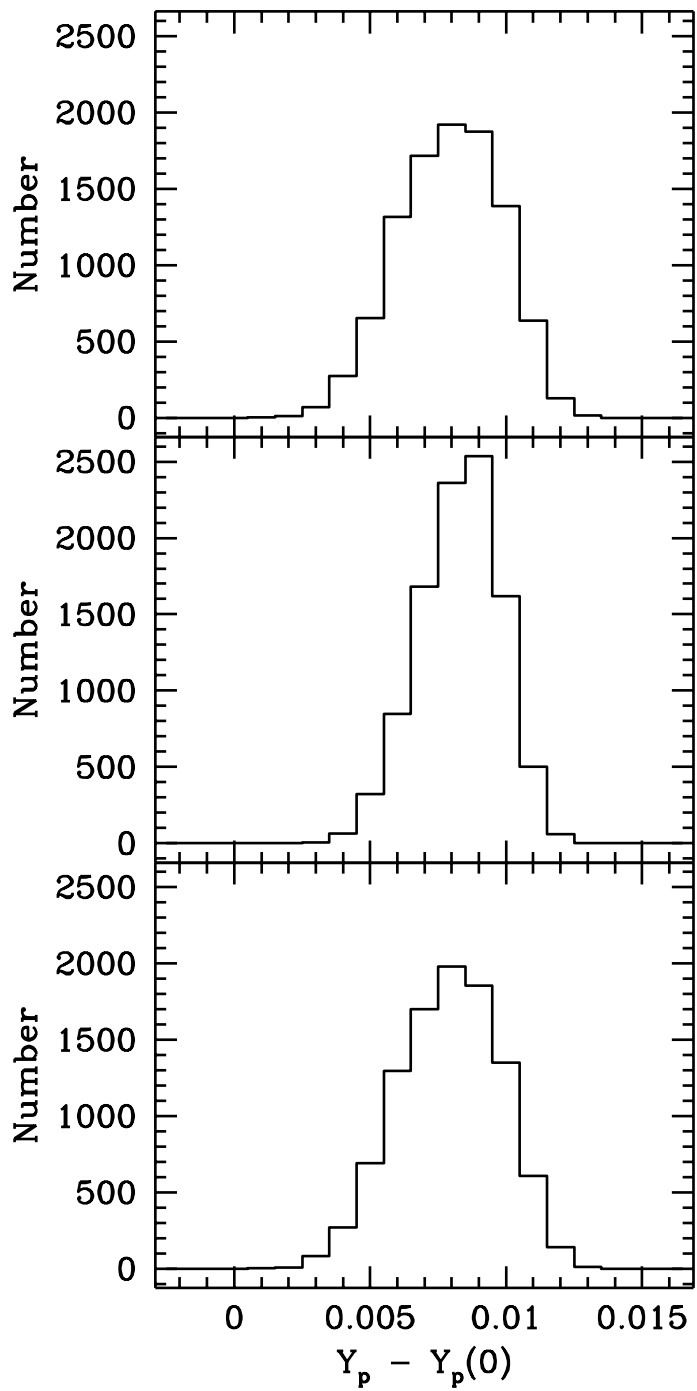
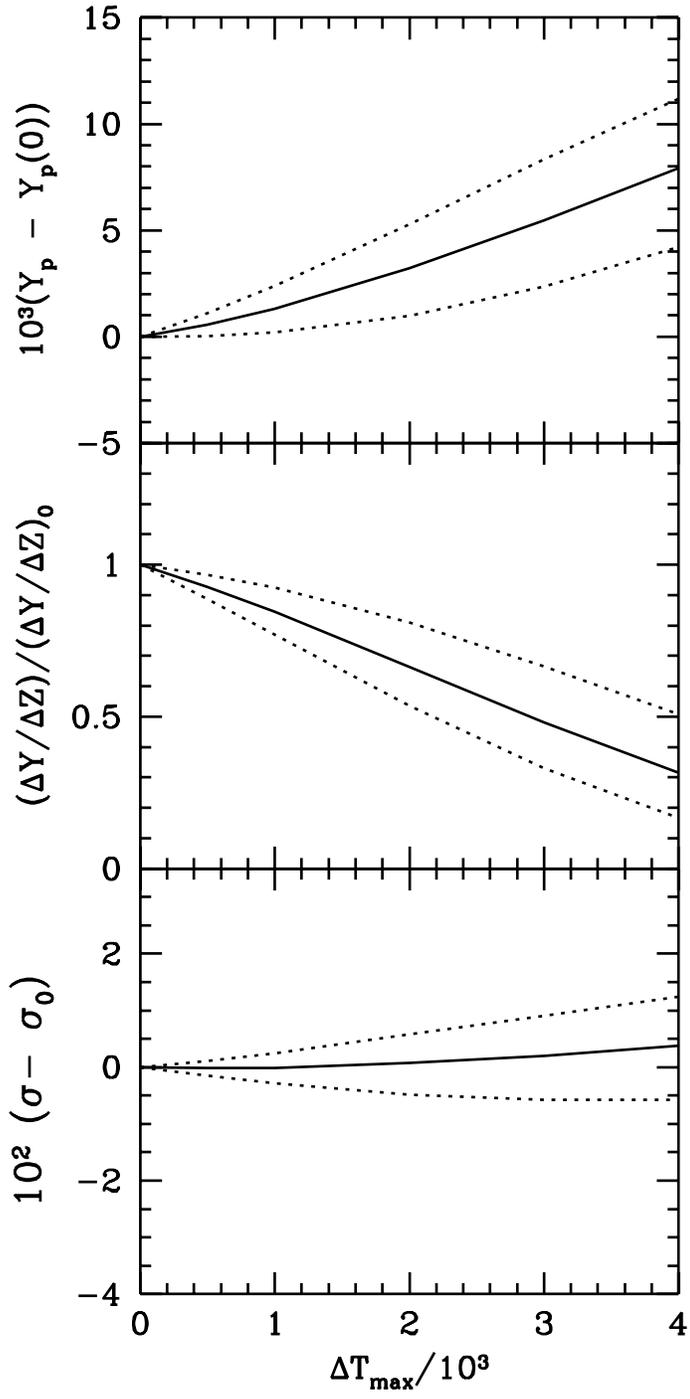
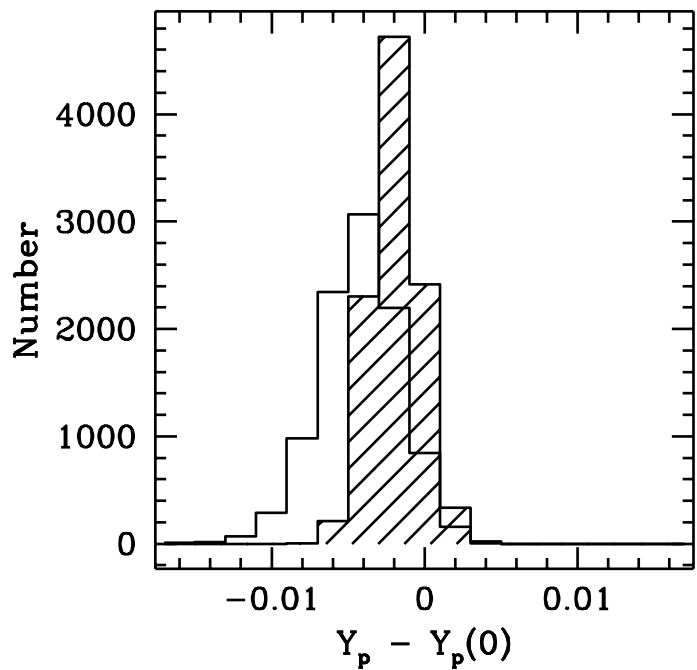
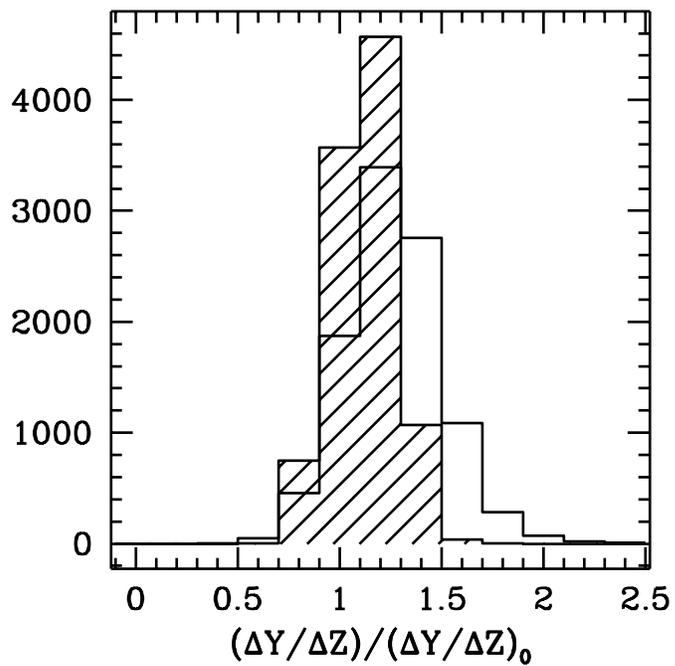
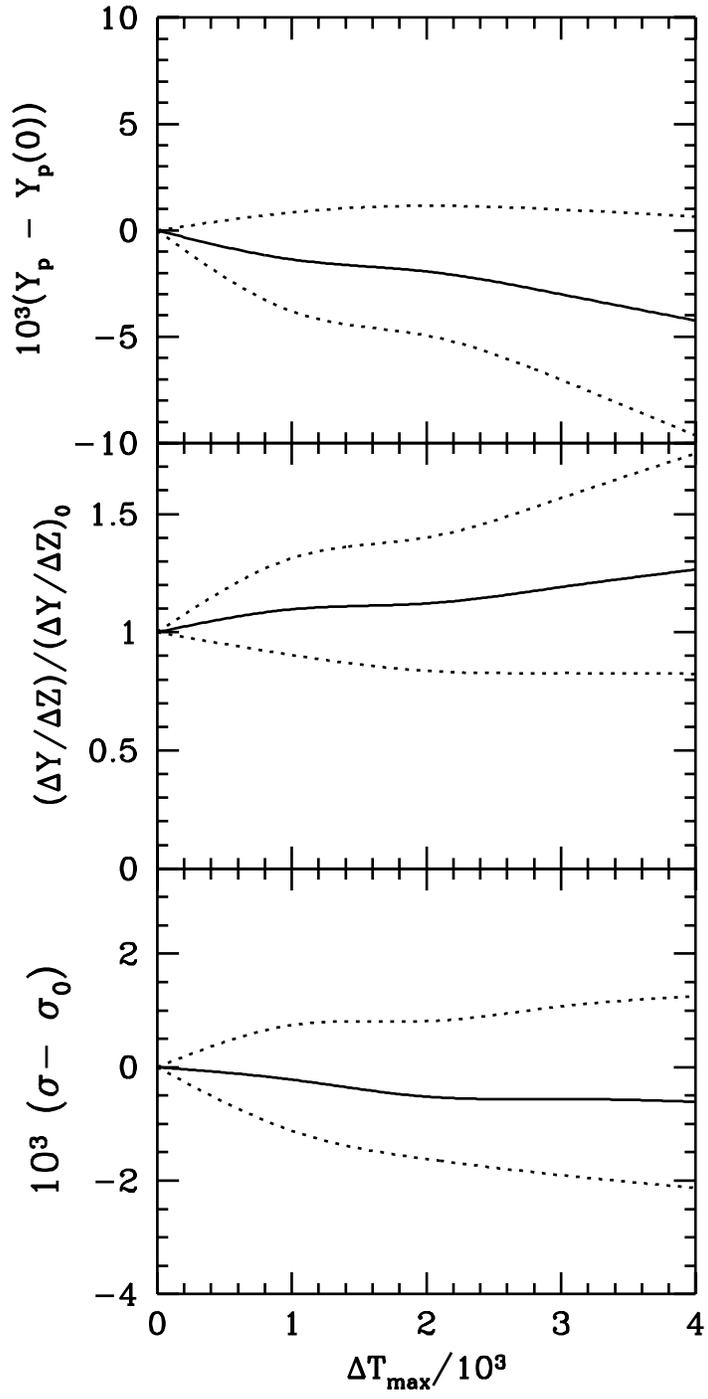
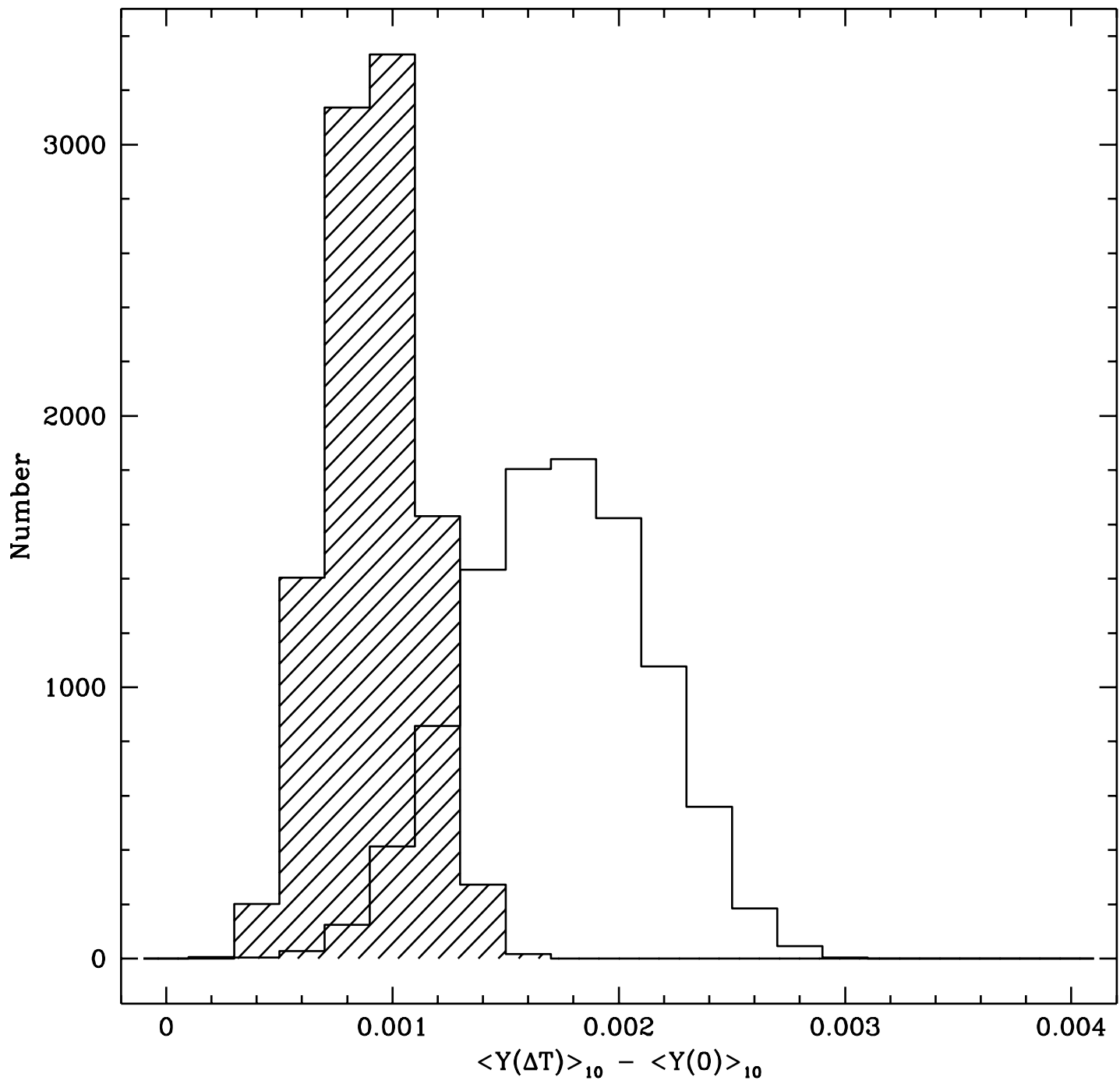
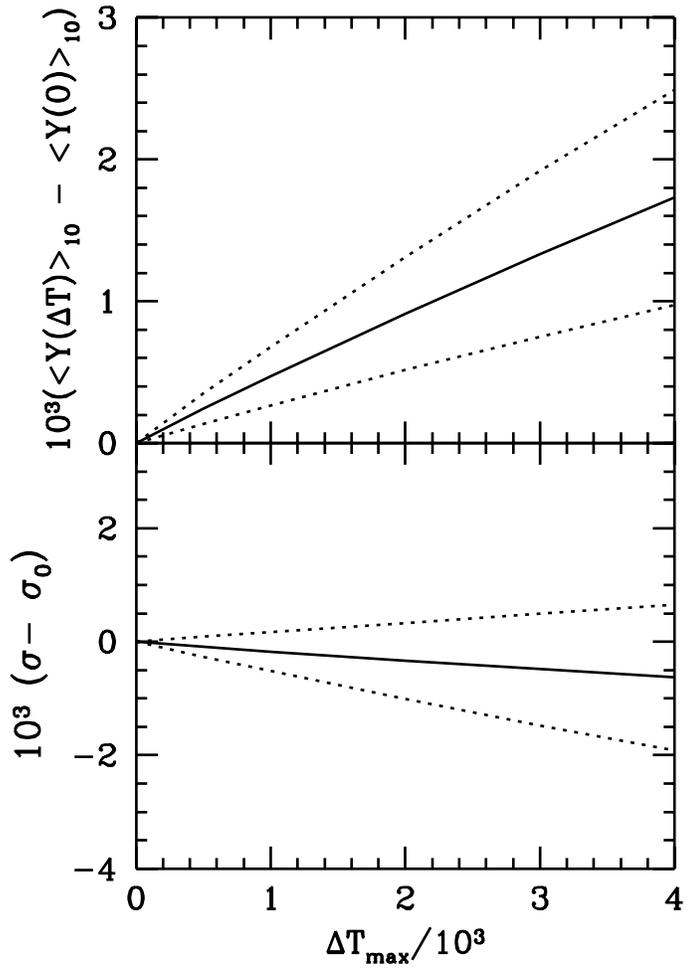
\end{document}